\documentstyle[11pt,IAUS215,twoside,epsf]{article}

\begin{document}

\title{The Analysis of the Rotating Disk of some Southern 
Galactic B[e] Stars}
\author{Marcelo Borges Fernandes, H.J.G.L.M. Lamers, Michaela Kraus}
\affil{Astronomical Institute, Utrecht University, P.O. Box 80000, NL-3508, TA, 
Utrecht, The Netherlands}
\author{Francisco X. de Ara\'ujo}
\affil{Observat\'orio Nacional, Rua General Jos\'e Cristino, 77, CEP: 20921-400, 
S\~ao Crist\'ov\~ao, Rio de Janeiro, Brasil}

\begin{abstract}
The spectra of stars with the B[e] phenomenon are 
dominated by features that are 
related to physical conditions of circumstellar material around these objects 
and are not intrinsic to the stars. Previous studies have shown that emission 
lines present in the optical spectra of these objects are formed in an 
equatorial rotating disk. The analysis of high and low resolution spectra, 
obtained by us with 1.52 
telescope in ESO for some Southern Galactic 
B[e] Stars, can give us information about the structure and velocity of the 
disk. We will describe the analysis of the unclassified B[e] star Hen 2-90.
\end{abstract}

\section{Introduction}

From a sample of stars with the B[e] phenomenon (Lamers et al. 1998) observed by 
us with the FEROS and B\&C spectrograph at 1.52 telescope in ESO (La Silla, 
Chile - agreement ESO/ON), we have analyzed the circumstellar medium around 
{\index Hen 2-90}, a unclassified B[e] star (that is sometimes classified in the 
literature as a compact planetary nebulae) considering the presence of a 
rotating disk. Its presence is suggested by images taken with WFPC2 in HST 
(Sahai \& Nyman 2000) and also by the presence of double peaked profiles in its 
spectrum. Our analysis is based on the comparison between the observed 
luminosity of [S\,{\sc ii}] lines and H$\alpha$ with those predicted by a model 
that will be described below.

\section{Double Peaks}

The existence of a rotating disk at Hen 2-90 is confirmed by the 
presence of many forbidden and permitted lines with double peaks (see figure 1). 
From the separation between two peaks we 
have found a $v$ $sin$ $i$ $\sim$ 20 km s$^{-1}$.

\vskip 60truecm
\vbox{\includegraphics{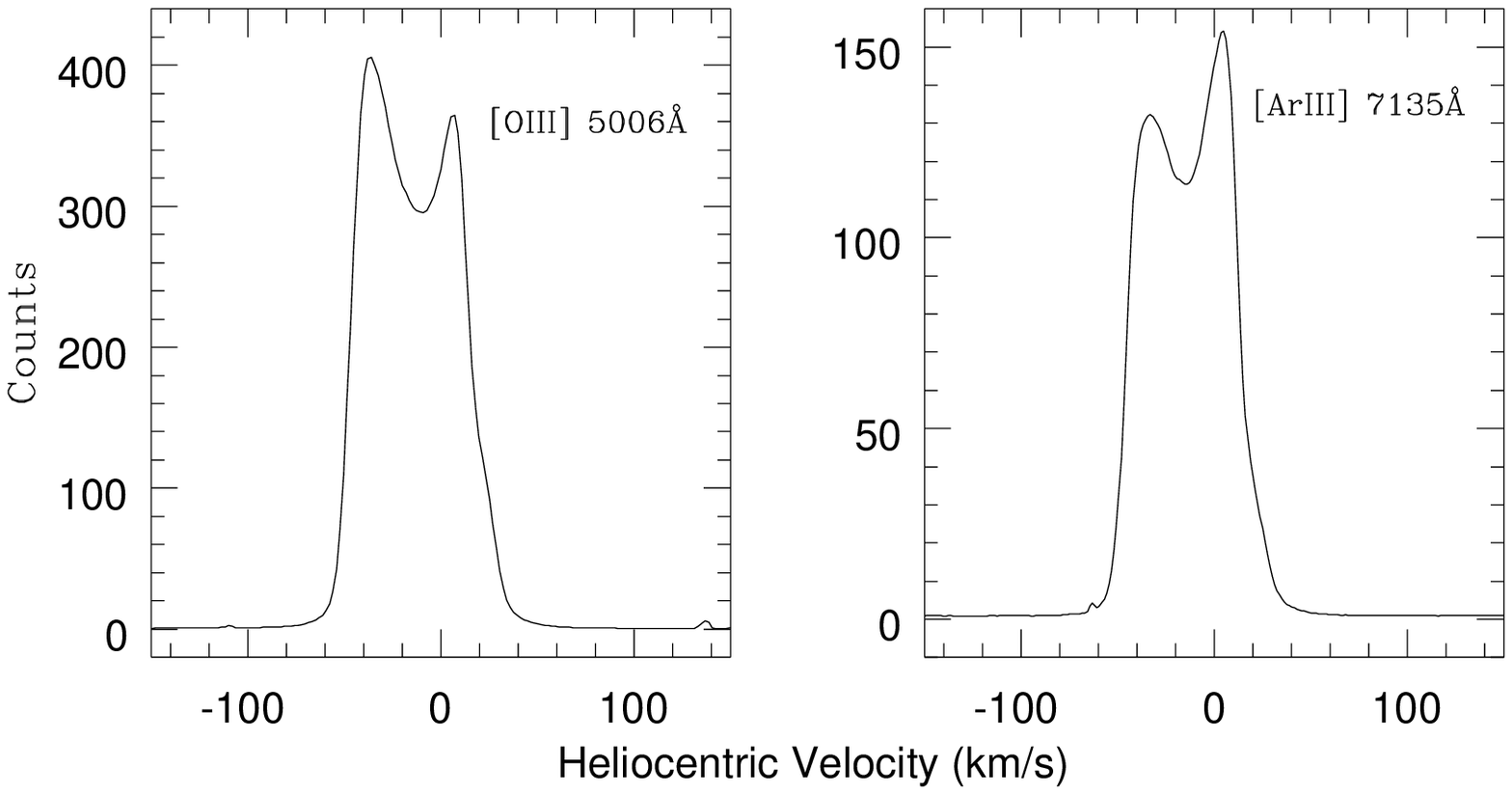}}
\vskip 5.5truecm
\noindent {\footnotesize {\bf \it Figure 1 -}[O\,{\sc iii}] and [Ar\,{\sc iii}] 
double peaked profiles present in 
the high resolution spectra (FEROS) of Hen 2-90. }

\section{Wind plus a disk: Our model}

\begin{itemize}
   
\item 2 spherical envelopes, considering one representing the polar wind and the 
other representing the disk. The ionization is calculated with a photoionization 
model for H, He and S. One envelope has a higher density than the other and we 
only use the equatorial sector with a width of 15 degrees. The other has a lower 
density and we are considering the luminosity that comes from the remaining 
region, outside the disk.

\item $\dot{M}$ is an input parameter

\item The [S\,{\sc ii}] lines come only from the disk and H$\alpha$ has two 
components that are coming from both regions.

\end{itemize}

\section{Results and Conclusions}

We have found a $\dot{M}$ $\sim$ 10$^{-9}$ M$_\odot$ year$^{-1}$ for the disk 
region, if all S is S\,{\sc ii}. This value is low compared with other compact 
planetary nebula. If the S\,{\sc ii} fraction (S\,{\sc ii}/S) is only 10$^{-3}$, 
then the 
$\dot{M}$ is $\sim$ 10$^{-7}$ M$_\odot$ year$^{-1}$, because we have found that 
$\dot{M}$ is proportional to
(S\,{\sc ii}/S)$^{-0.67}$ (Borges Fernandes et al. 2003).

We will apply this model for other stars in our sample. For this, we will 
improve the model including other elements like O and N.


\begin{references}
\reference Borges Fernandes, M., Lamers, H.J.G.L.M., Kraus, M. \& de 
Ara\'ujo, F.X. 2003, in preparation
\reference Lamers, H.J.G.L.M., Zickgraf, F.-J., de Winter, D., 
Houziaux, L. \& Zorec, J. 1998, A\&A 340, 117
\reference Sahai, R., \& Nyman, L.-A. 2000, ApJ 537, L145       
\end{references}
\end{document}